\documentclass[a4paper]{article}
\usepackage{INTERSPEECH2019}
\usepackage[bottom]{footmisc}
\usepackage{epsfig,amssymb,amsmath,rotating}
\ninept
\usepackage{pgfplots}
\usepackage{grffile}
\usepackage{float}
\PassOptionsToPackage{hyphens}{url}
\usepackage[bookmarks=false,hidelinks]{hyperref}

\usepackage[caption=false,font=footnotesize,subrefformat=parens,labelformat=parens]{subfig}
\usepackage{tikz}
\usepackage{pgfplots}
\usepackage{grffile}
\usepackage{float}
\pgfplotsset{compat=newest}
\makeatletter
\g@addto@macro\@floatboxreset\centering
\makeatother
\def\addlegendimage{\csname pgfplots@addlegendimage\endcsname}
\usepackage{graphicx}
\usepackage{cmap}
\usetikzlibrary{shapes.misc}
\usetikzlibrary{mindmap}
\usetikzlibrary{plotmarks}
\usetikzlibrary{arrows.meta}
\usepgfplotslibrary{patchplots}

\tikzset{
    font=\scriptsize
}

\newcommand{\setvariable}[2]{
	\let#1\relax
	\newcommand{#1}{#2}
}

\renewcommand{\vec}[1]{\bm{#1}}

\definecolor{darkgreen}{rgb}{0.2,0.7,0.3}
\definecolor{darkorange}{rgb}{1,0.55,0}

\pgfplotsset{
    mated/.style={ybar interval, fill opacity=0.6, area legend, 
        fill=darkgreen, draw=white},
    nonmated/.style={ybar interval, fill opacity=0.6, area legend, 
        fill=white!30!red, draw=white},
    histAxis/.style={
        width=\figwidth,
        height=\figheight,
        scale only axis,
        ymin=0,ymax=1,
        ylabel={Rel. freq.},
        xmajorgrids,ymajorgrids,},
    rawHistAxis/.style={histAxis,xlabel={Scores},xmin=-20,xmax=40},
    llrHistAxis/.style={histAxis,xlabel={Ideal LLRs},xmin=-10,xmax=10},
}

\usepackage{eurosym}

\setvariable{\colorScalePurple}{PiYG-11-1}
\setvariable{\colorScaleOrange}{Oranges-9-4}
\setvariable{\colorScaleBlue}{Spectral-5-5}
\setvariable{\colorScaleRed}{hda}
\setvariable{\colorScaleGreen}{Spectral-5-4!80!black}

\usepackage[justification=centering]{caption}
\usepackage[labelsep=colon]{caption}
\usepackage{amsmath}
\usepackage{amsfonts}
\usepackage{amssymb}
\usepackage{url}
\usepackage{graphicx}
\usepackage{epstopdf}
\usepackage{mwe}
\usepackage{multirow}
\usepackage[export]{adjustbox}
\usepackage{tikz}
\usepackage{algorithm}
\usepackage{kantlipsum}
\usepackage{xpatch,xcolor}
\usepackage{graphicx}
\usepackage{afterpage}
\usepackage{balance}
\usepackage{xcolor}
\tikzset{
    font=\scriptsize
}

\usepackage{balance}

\title{Spoofing Attack Detection using the Non-linear Fusion of Sub-band Classifiers}
\name{Hemlata Tak, Jose Patino, Andreas Nautsch, Nicholas Evans and Massimiliano Todisco}

  \address{EURECOM, Sophia Antipolis, France}
\email{lastname@eurecom.fr}

\begin{document}

\maketitle
\begin{abstract}
 
The threat of spoofing can pose a risk to the reliability of automatic speaker verification.  Results from the bi-annual ASVspoof evaluations show that effective countermeasures demand front-ends designed specifically for the detection of spoofing artefacts.  Given the diversity in spoofing attacks, ensemble methods are particularly effective.  The work in this paper shows that a bank of very simple classifiers, each with a front-end tuned to the detection of different spoofing attacks and combined at the score level through non-linear fusion, can deliver superior performance than more sophisticated ensemble solutions that rely upon complex neural network architectures.  Our comparatively simple approach outperforms all but 2 of the 48 systems submitted to the logical access condition of the most recent ASVspoof 2019 challenge.
\end{abstract}

\section{Introduction}

A great deal of research in ASV anti-spoofing has focused on the design of specific front-ends  tuned to capture artefacts that characterise manipulated or synthetic speech.  The results of the ASVspoof 2019 challenge also show that reliable performance usually demands the fusion of scores derived from an ensemble of different front-ends.  This observation suggests that no single front-end can detect reliably the full range of artefacts produced by different spoofing attacks.

There is evidence that spoofing artefacts lie at the sub-band level~\cite{Sahidullah15, sriskandaraja2016investigation,witkowski2017audio,nagarsheth2017replay,lin2018replay,yang2019significance,garg2019subband} and that these can only be detected reliably using front-ends that have high spectral resolutions in the same bands~\cite{jung2019replay,odyssey2020CQCC}.  This means that conventional cepstral processing may be detrimental to anti-spoofing performance in the sense that cepstral analysis averages information across the full spectrum, rather than emphasising information at the sub-band level.  This in turn may explain why reliable performance is obtained only through the fusion of several systems, with similar performance not being achieved with single systems.  Results from the ASVspoof 2019 challenge support this hypothesis.  Although there are likely to be additional contributing factors, the top-performing fused-system submission achieved an equal error rate (EER) of 0.22~\% whereas the same team's single-system submission achieved an EER of 11.40~\%, some 52 times higher.

The observation that different front-ends are required to detect artefacts located within different sub-bands may mean that the usual approaches to fusion will be sub-optimal.  This is principally because any single spoofing attack may only be detected reliably by a single countermeasure within an ensemble.  In this case, linear approaches to score fusion may not exploit the complementarity of each countermeasure to their full potential; linear combinations of mostly non-informative scores may serve to dilute informative scores. Non-linear approaches to fusion may hence be better suited to such scenarios.

The work reported in this paper was designed to test these hypotheses, namely that: (i)~an ensemble of relatively simple countermeasures, each tuned to the detection of artefacts in different sub-bands, may help to improve spoofing detection performance beyond what can be achieved through the fusion of different countermeasures operating at the full-band level; (ii)~non-linear fusion may better exploit complementarity beyond what can be achieved with linear approaches.  To the best of our knowledge, while some work has already demonstrated the benefit of ensemble methods, e.g.~\cite{ji2017ensemble,chettri2019ensemble,Ensemble2020}, none of the past work has investigated the reasons {\it why} they are beneficial, and neither have they explored ensemble methods in the context of fused, attack-optimised, sub-band front-ends.

\section{Research hypotheses}

\begin{figure}
	\centering
	\includegraphics[width=8cm,trim={0 0.2cm 0 0},clip]{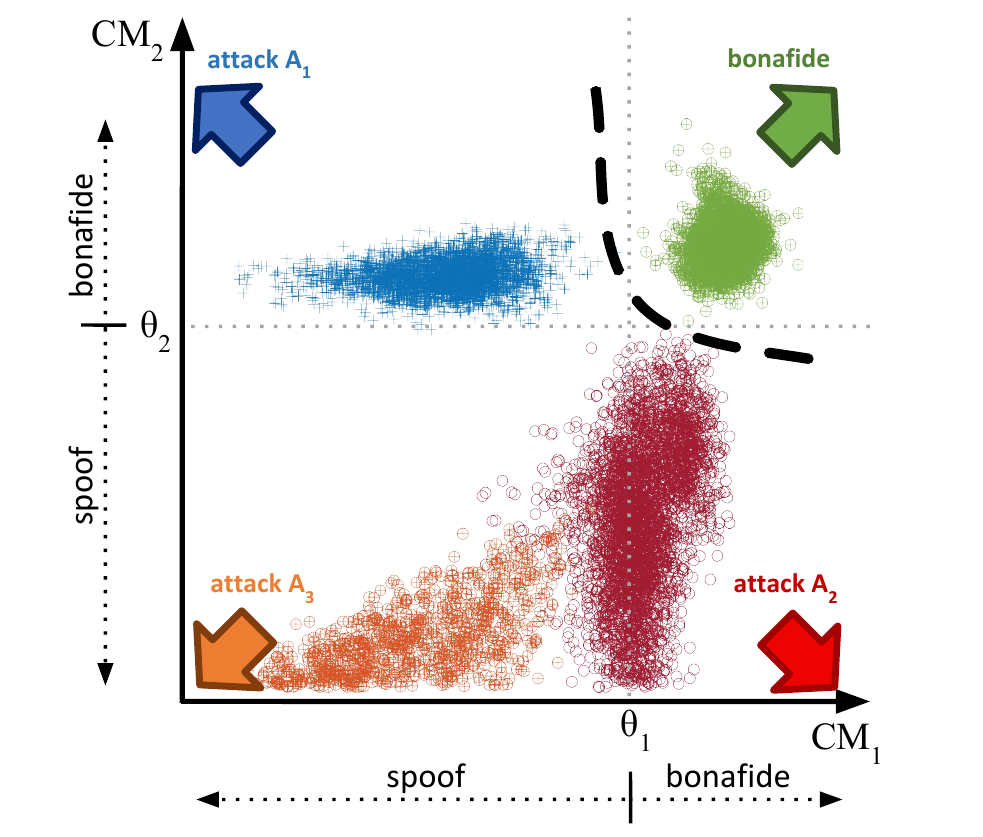} 	
	\caption{A scatter plot of scores for countermeasures $\text{CM}_1$ and $\text{CM}_2$.  Clusters correspond to bona fide utterances (green) and three spoofing attacks (A1-blue, A2-red, A3-orange).  The dashed black line indicates a non-linear decision boundary that best separates bona fide from spoofed utterances.
	}
	\label{fig:fusionDiagram}
    \vspace{-0.2cm}
\end{figure}

To help illustrate the ideas explored in this work, we consider the hypothetical anti-spoofing example illustrated in
Fig.~\ref{fig:fusionDiagram}.  Plotted on each axis are the scores produced by two different spoofing countermeasures: $\text{CM}_1$ and~$\text{CM}_2$.  Countermeasure $\text{CM}_1$ is tuned to detect artefacts present within a lower sub-band, at 0-4~kHz for example.  Countermeasure $\text{CM}_2$ is tuned to detect artefacts within a higher sub-band, at 4-8~kHz for example.  Each point in the plot signifies the scores produced by each countermeasure for a set of utterances.  Scores for bona fide (genuine / not spoofed) utterances are illustrated by green points (top-right). Also shown are scores for three different types of spoofing attacks: attack $\text{A}_1$,  characterised by artefacts predominantly at low frequencies (blue points, top-left); attack $\text{A}_2$, characterised by artefacts at high frequencies (red points, bottom right); attack $\text{A}_3$, which exhibits artefacts at both low and high frequencies (orange points, bottom left).  The first hypothesis under investigation in this paper is that different spoofing attacks are characterised by artefacts within different sub-bands and that an ensemble of different front-ends are needed in order to detect such artefacts reliably.

Both countermeasures produce predominantly high scores for bona fide utterances; as per standard ASVspoof practice, high countermeasure scores reflect bona fide trials, whereas low scores reflect spoof trials. Since $\text{CM}_1$ and $\text{CM}_2$ and their respective thresholds $\theta_1$ and $\theta_2$ are tuned for the detection of spoofing attacks $A_1$ and $A_2$ respectively, spoofing attack $\text{A}_1$ provokes mostly low scores for $\text{CM}_1$ and mostly high scores for $\text{CM}_2$, and vice versa for attack $\text{A}_2$.  Attack $\text{A}_3$ provokes low scores for both countermeasures. Considering \emph{multiple diverse attacks and countermeasures}, a notional decision boundary that best separates bona fide from spoofing utterances might correspond to a non-linear function.  Linear score fusion operators may not perform well in this case, leading to poor reliability.  The second hypothesis under investigation in this paper is that a non-linear approach to score fusion or system combination is needed in order to best exploit the complimentarity of an ensemble of countermeasures tuned for the detection of specific spoofing attacks.

\section{Experimental setup}

Experiments were performed with the logical access (LA) partition of the ASVspoof 2019 database and standard protocols~\cite{todisco2019asvspoof,wang2019asvspoof}.  Sub-band analysis and fusion experiments are furthermore based upon one of the two standard baselines.  The database, baseline and assessment metric are all described here.

\subsection{Database and metrics}

The ASVspoof 2019 LA database consists of three independent partitions: train; development; evaluation. 
Spoofed speech in each dataset is generated using a set of different  
speech synthesis and voice conversion algorithms~\cite{wang2019asvspoof}. 
There is a total of 19 different spoofing attacks. 
Attacks in the training and development set were created with a set of 6 different algorithms (A01-A06), whereas those in the evaluation set were created with a set of 13 algorithms (A07-A19). 
The three partitions are used according to standard ASVspoof practice~\cite{ASVspoof19_evalplan}.

The primary metric used in this work is the minimum  normalised tandem detection cost function (t-DCF) metric~\cite{kinnunen2018t,kinnunen-tDCF-TASLP}.
The t-DCF reflects the impact of spoofing and countermeasures (CMs) upon the reliability of an automatic speaker verification (ASV) system.  
To give a more intuitive impression of countermeasure performance, results are also reported in terms of the pooled equal error rate (EER) computed using~\cite{brummer2013bosaris}.

\subsection{Baselines}
\label{sec:baselines}

The ASVspoof 2019 baseline systems use either constant Q cepstral coefficient (CQCC)~\cite{todisco2016new,todisco2017constant} or linear frequency cepstral coefficient (LFCC)~\cite{Sahidullah15} front-ends and a common Gaussian mixture model (GMM) back-end. Both LFCC-GMM and CQCC-GMM baseline systems are described in full in~\cite{todisco2019asvspoof,ASVspoof19_evalplan}.
While results for both baselines are reported in this paper, and while most of our group's recent work is based upon constant Q transform (CQT)   representations~\cite{brown1991calculation}, for reasons discussed later, experimental work reported here is all based upon modifications to the LFCC front-end. The baseline LFCC front-end configuration includes 20ms frame-blocking with 10ms shift, a filterbank with $20$ linearly-scaled filters and 20 static, velocity ($\Delta$) and acceleration ($\Delta \Delta$) coefficients, thereby giving 60-dimensional feature vectors. 

The GMM back-end classifier has two 512-component Gaussian models.  The first is a model of bona fide speech whereas the second is a model of spoofed speech, with both being learned using bona fide and spoofed speech data from the ASVspoof 2019 LA training partition. Scores are log-likelihood ratios (LLRs) computed in the usual way.

\section{Sub-band front-ends}
Our recent work~\cite{odyssey2020CQCC} showed that spoofing artefacts reside at the sub-band level and that these are best detected with front-ends that exhibit a high spectral resolution within the same sub-band.  That work used two different CQCC front-ends that were tuned to increase the spectral resolution at either low or high frequencies.  The current work extends this idea by considering an ensemble of classifiers with each one being tuned for the detection of a set of specific spoofing attacks and the associated artefacts \emph{no matter where they are in the spectrum.}  Since the CQT has a non-linear spectral resolution~\cite{brown1991calculation,schorkhuber2014matlab} which is difficult to tune to specific sub-bands, the work was performed by adapting the baseline LFCC front-end described in Section~\ref{sec:baselines}.

Described here is the strategy of spectral resolution and front-end tuning at the sub-band level. Also presented are results for each front-end when used with a GMM back-end and tested against each spoofing attack in the ASVspoof 2019 LA database.

\subsection{Spectral resolution}

\label{optimized_lfcc}

Use of a spectral resolution that is \emph{too} high will result in noisy features.  Hence, before sub-band optimisation, we set out first to optimise the spectral resolution at the full-band level.  This work was performed using the the full ASVspoof 2019 LA training and development subsets. 

While other techniques could also have been applied, e.g.~zero padding, we simply modified the baseline LFCC front-end (Sec.~\ref{sec:baselines}) to use a 30~ms window with a 15~ms shift and used a 1024-point Fourier transform.  The resolution was then decreased using a filterbank in the usual fashion with a number of filters $N$~\cite{deller1993discrete}.  For any one experiment, training and development data were processed with the given front-end before the GMM back-end was re-learned and used to process the development data in otherwise identical fashion to the baseline.

Results depicted in Table~\ref{Tab:No. of filters} show CM performance in terms of the min t-DCF and EER against the number $N$ of filterbank filters (first 3 columns).  For $N>30$ filters, both the min t-DCF and the EER are zero.  An alternative approach to optimisation is hence necessary.  We elected arbitrarily to use the Bhattacharyya distance~\cite{bhattacharyya1943measure} between the CM score distributions for bona fide and spoofed trials given by:

\begin{equation}
D_{B}(b,s)=\frac{1}{4}ln \left( \frac{1}{4}\left(\frac{\sigma_{b}^{2}}{\sigma_{s}^{2}}+\frac{\sigma_{s}^{2}}{\sigma_{b}^{2}}+2\right)\right)+\frac{1}{4}\left(\frac{(\mu_{b}-\mu_{s})^{2}}{\sigma_{b}^{2}+\sigma_{s}^{2}}\right),
\nonumber
\label{eq:Distence Measure}
\end{equation}

\noindent where subscripts $b$ and $s$ indicate parameters for bona fide and spoofed score distributions and where $\mu$ and $\sigma$ refer to the  means and standard deviations respectively.  Results in the last column of Table~\ref{Tab:No. of filters} show that the distance between score distributions increases for $N>30$ filters, but with little gain beyond $N=70$ filters, which is the configuration used for all further experiments reported in this paper.

\begin{table}[!t]
	\centering
	\caption{min t-DCF, EER and Bhattacharyya distance between bona fide and spoofed score distributions for different numbers of subband filters $N$.  Baseline configuration illustrated in bold; selected configuration in italics.}
	\setlength\tabcolsep{2pt}
	\begin{tabular}{ *{4}{c}}
		\hline
Filters ($N$)& min t-DCF&EER (\%)&$D_B$ \\
\hline\hline
{\bfseries 20} & {\bfseries 0.2110} & {\bfseries 2.71} & {\bfseries 0.1338}\\
\hline
30&0.0000&0.79&0.1706\\
\hline
40&0.0000&0.00&0.1770\\
\hline
50&0.0000&0.00&0.1785\\
\hline
60&0.0000&0.00&0.1793\\
\hline
\textit{70}&\textit{0.0000}&\textit{0.00}&\textit{0.1826}\\
\hline
80&0.0000&0.00&0.1788\\
\hline
90&0.0000&0.00&0.1823\\
\hline
100&0.0000&0.00&0.1830\\
\hline
120&0.0000&0.00&0.1820\\
\hline
\end{tabular}
\label{Tab:No. of filters} 
\vspace{-0.2cm}
\end{table}

\subsection{Centre of Mass Function}
\label{Section:CMF}

\begin{figure}[!t]
	\centering
%
%
\begin{tikzpicture}
\begin{axis}[%
width=3.4cm,
height=3.4cm,
scale only axis,
point meta min=0,
point meta max=1,
axis on top,
xmin=0.5,
xmax=2000.5,
xtick={0.5,500,1000,1500,2000},
xticklabels={{0},{2},{4},{6},{8}},
xlabel style={font=\bfseries\color{white!15!black}},
xlabel={$\text{f}_{\text{min}}$} {\text{(kHz)}},
y dir=reverse,
ymin=0.5,
ymax=2000.5,
ytick={0.5,500,1000,1500},
yticklabels={{8},{6},{4},{2}},
ylabel style={font=\bfseries\color{white!15!black}},
ylabel={$\text{f}_{\text{max}}$} {\text{(kHz)}},
axis background/.style={fill=white},
axis x line*=bottom,
axis y line*=left,
legend style={legend cell align=left, align=left, draw=white!15!black},
colormap/jet,
colorbar,
colorbar style={ylabel style={font=\bfseries\color{white!15!black}}, ylabel={min t-DCF}}
]
\addplot [forget plot] graphics [xmin=0.5, xmax=2000.5, ymin=0.5, ymax=2000.5] {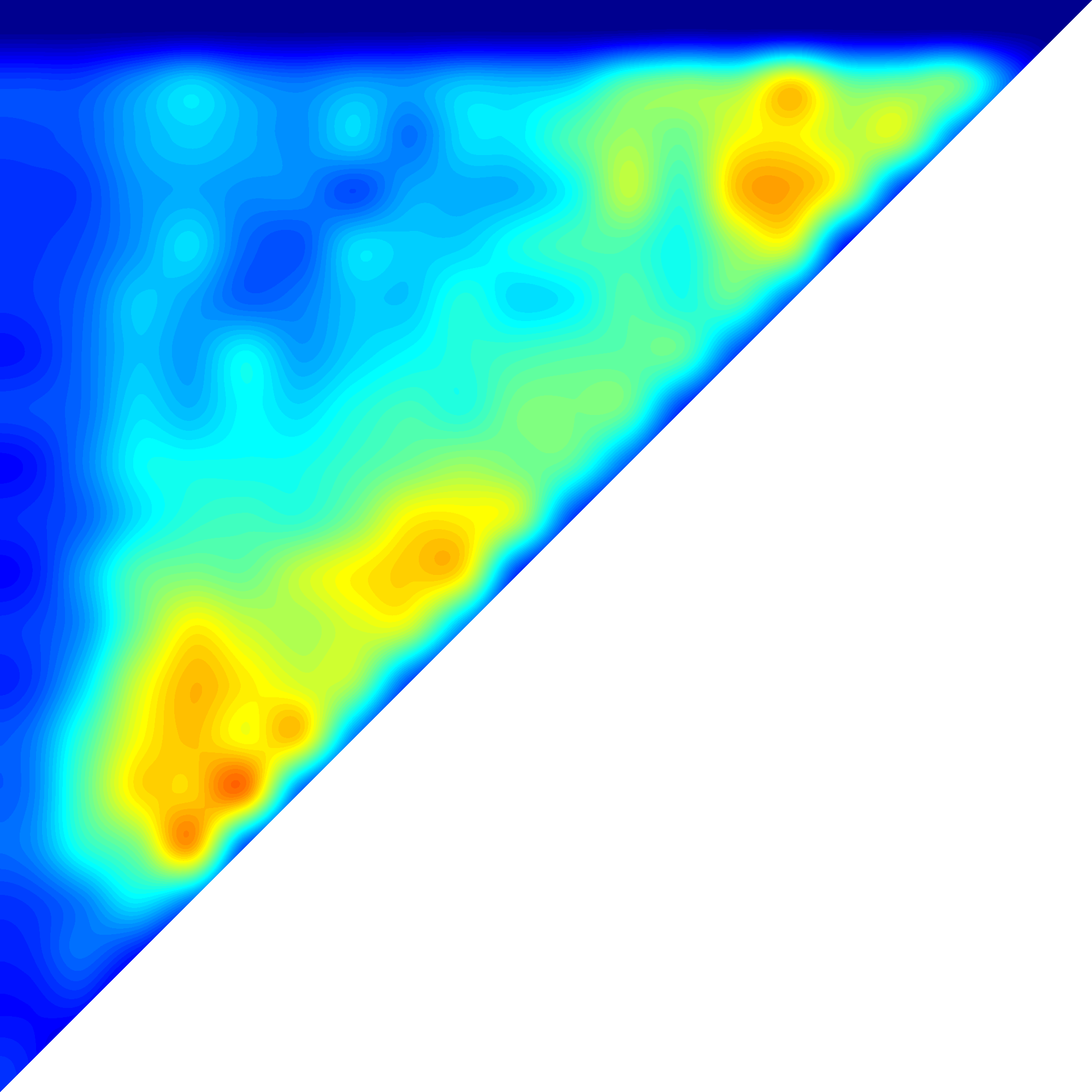};
\addplot [white,ultra thick,mark=x,mark size=.5em] coordinates{(750,80)};
\end{axis}
\end{tikzpicture}%
	\caption{A 2-D heatmap visualisation (see~\cite{odyssey2020CQCC}) illustrating sub-band level CM performance for attack A04 of the ASVspoof 2019 LA database. 
	The cut-in frequencies $f_{\text{min}}$ and cut-off frequencies $f_{\text{max}}$ are indicated on horizontal and vertical axes respectively. Those of the CoM-defined sub-band is indicated by the white cross.}

	\label{Figure. triangle explaination}
\vspace{-0.3cm}
\end{figure}

Attack-specific, sub-band front-ends are designed using heat-map visualisations~\cite{odyssey2020CQCC} which show CM performance at the sub-band level.  An example for the A04 attack is illustrated in Fig.~\ref{Figure. triangle explaination}. The heat-map colour signifies CM performance in terms of t-DCF for a front-end with cut-in and cut-off frequencies of $f_{\text{min}}$ (x-axis) and $f_{\text{max}}$ (y-axis) respectively. We used the centre-of-mass (CoM) approach~\cite{Feynman:1494701} to identify a single point in the heat-map and hence to define a specific sub-band for the detection of each attack in the development subset (A01--A06). 

The CoM is a crude means of coping with a noisy surface containing multiple minima. The CoM of a distribution of mass in space is the unique point where the weighted relative position of the distributed mass sums to zero. 
We consider the 2D heat-map as a system of particles $P _ { i }$ where $i = 1 , \ldots , n$.  Each particle has coordinates  $r _ { i } = [f_{\text{min}}^{i},f_{\text{max}}^{i}]$ and mass $m _ { i } = ($min t-DCF$ _ { i }) ^ {-1}$. The coordinates $R = [f_{\text{min}}^{CoM},f_{\text{max}}^{CoM}]$ of the CoM satisfy the condition $\sum _ { i = 1 } ^ { n } m _ { i } ( r _ { i } - R ) = 0$. Solving for $R$ yields: 
\begin{equation}
R = \frac { 1 } { M } \sum _ { i = 1 } ^ { n } m _ { i } r _ { i }
\label{eq:CoM}
\end{equation}
\noindent where $M$ is the sum of the masses of all the particles in the full heat-map.
We obtain a different $R$ for each attack and hence define six attack-optimised, sub-band CMs.  For attack A04 the CoM point illustrated by the white cross in Fig.~\ref{Figure. triangle explaination} signifies a sub-band of 3209 to 8000~Hz.  The CoM-defined sub-bands for each attack A01-06 are listed in the first column of Table~\ref{Tab: results on dev and eval}.

\subsection{Sub-band CM results}

Results presented in Table \ref{Tab: results on dev and eval} show performance in terms of t-DCF for the six sub-band and one full-band CM. The bandwidth of each system is illustrated in the first column.  

Results for the development set (columns A01-06) show that sub-band CMs all yield zero t-DCFs for the attacks for which they are designed (results in boldface), as they also do for some other attacks.  This is not surprising since there is some considerable spectrum overlap in the set of sub-band CMs.  Interestingly, the full-band CM is the only one to achieve zero t-DCF for all six attacks in the development set.  Results for the evaluation set (columns A07-19) show that the full-band CM gives similar or substantially lower t-DCFs than sub-band CMs.  These observations are confirmed by pooled t-DCFs (columns P1) for both the development and evaluation subsets.  The questions now are: (i)~whether or not the fusion of attack-specific, sub-band CM scores can give better performance even when their individual performance is poor relative to the full-band CM; (ii)~what should be the fusion mechanism.

\begin{table*}[!t]
	\centering
	
	\caption{Results in terms of min~t-DCF for development (A01-A06) and evaluation (A07-A19) partitions and respective pooled min~t-DCF (P1) and pooled EER (P2). Results in boldface signify the attack for which each sub-band is optimised, e.g.\ the CM designed for attack A01 operates within a sub-band of 2011 to 6403~Hz.}
	\setlength\tabcolsep{1.6pt}
	\begin{tabular}{c || *{6}{c} |cc|| *{13}{c}| cc} 
		\hline
	
		Freq-bands &A01&A02&A03&A04&A05&A06&P1&P2&A07&A08&A09&A10&A11&A12&A13&A14&A15&A16&A17&A18 &A19& P1&P2  	\\ 
		\hline\hline
		2011-6403&\textbf{0.00}&0.00&0.00&0.22&0.25&0.79&0.25&0.11&0.37&0.03&0.00&0.55&0.03&0.18&0.31&0.17&0.15&0.25&0.41&0.60&0.93	&	0.34&	13.28\\
		\hline
	    2410-5604&0.00&\textbf{0.00}&0.00&0.31&0.42&0.91&0.33&0.13&0.35&0.06&0.00&0.54&0.08&0.23&0.31&0.16&0.16&0.33&0.58&0.87&0.99  &	0.39&	15.50\\
		\hline
		2011-5604&0.00&0.00&\textbf{0.00}&0.27&0.31&0.82&0.29&0.12&0.37&0.06&0.00&0.56&0.08&0.22&0.28&0.18&0.17&0.30&0.48&0.81&0.99  &0.38&	14.75\\
		\hline
	3209-8000&0.00&0.00&0.00&\textbf{0.00}&0.15&0.00&0.03&0.01&0.00&0.00&0.00&0.54&0.00&0.41&0.71&0.10&0.23&0.00&0.47&0.29&0.00      &	0.26&	10.59\\
		\hline
		15.62-4806&0.00&0.00&0.00&0.16&\textbf{0.00}&0.51&0.18&0.08& 0.45&0.01&0.00&0.54&0.07&0.09&0.09&0.12&0.13&0.16&0.33&0.60&0.85     &	0.31&	12.31\\
		\hline
		3608-8000&0.00&0.00&0.00&0.00&0.12&\textbf{0.00}&0.04&0.01& 0.00&0.00&0.00&0.55&0.00&0.40&0.79&0.09&0.31&0.00&0.55&0.24&0.00         &  	0.27&	11.55\\
		\hline
		full-band&0.00&0.00&0.00&0.00&0.00&0.00&0.00&0.00& 0.00&0.00&0.00&0.15&0.00&0.11&0.07&0.06&0.06&0.00&0.35&0.07&0.00 	&\textbf{0.09}&	\textbf{03.50}\\
		\hline
	\end{tabular}
	\label{Tab: results on dev and eval}

\end{table*}

\section{Fusion}

Fusion experiments aim to assess the second research hypothesis in this work, namely that a non-linear approach to score fusion or system combination is needed in order to best exploit the complimentarity of sub-band CMs.  We used four different fusion methods to obtain a single score from the set of seven scores: six sub-band CMs and one full-band CM.\\   

Approaches to fusion include: a support vector machine (SVM)~\cite{cortes1995support} with a seventh order polynomial kernel\footnote{Linear and residual basis function kernels were also tested and yielded inferior results.  These experiments are not reported here.}; multinomial logistic regression~\cite{bishop:2006:PRML}; traditional linear fusion~\cite{brummer2013bosaris}. 
Also tested was a GMM-based approach to fusion for which 64-component models are learned from the set of scores for bona fide and spoofed classes.  
This approach was used previously for the fusion of ASV and CM scores~\cite{todisco2018integrated}.
Both the GMM and SVM are non-linear approaches to fusion; the others are linear.  All but the SVM approach produce log-likelihood ratio outputs.

\subsection{Fusion results}
Fusion results for the four systems are shown in boldface in Table~\ref{Tab:comparsion results}.  With a t-DCF of 0.0740, the non-linear GMM approach gives the best performance.
The next best system is the non-linear SVM approach with a t-DCF of 0.0748.  The performance of the two linear approaches yield t-DCFs of 0.0911 and 0.1182.  These findings would seem to confirm the hypothesis that a non-linear approach is better suited to the fusion of sub-band CM scores.  This is because spoofing artefacts that are localised in the spectrum may be detected only by sub-band CMs whose focus is directed towards the same parts of the spectrum and hence be detected reliably by a sub-set of CMs only (or even only a single CM).  In this case, full-band CMs may dilute relevant information by smoothing across the spectrum and linear approaches to fusion may not identify the best decision boundary between bona fide and spoofed speech; such an optimal decision boundary might be non-linear.

\subsection{Performance comparison and discussion}
Table~\ref{Tab:comparsion results} also shows results for the two ASVspoof 2019 baseline systems (B1 and B2, last two rows) and the top-performing four (out of 48 submissions) challenge results~\cite{todisco2019asvspoof}.  The latter are signified by their anonymous ASVspoof 2019 identifiers T05, T45, T60 and T24~\cite{todisco2019asvspoof}.  Only T45~\cite{lavrentyeva2019stc} and T60~\cite{chettri2019ensemble} system details are in the public domain but from these and the description of the ASVspoof 2019 challenge in~\cite{todisco2019asvspoof}, it is known that all four of these competing systems are based upon an ensemble of comparatively complex neural network based architectures, as opposed to a simple GMM-based solution used in our work.  Furthermore, they used a combination of \emph{multiple, different} front-end parameterisations, unlike the use of the \emph{single, same} base front-end used in our work. While we acknowledge that this comparison is between evaluation and post-evaluation results, both non-linear GMM-based and SVM approaches to fusion outperform all but two of the 48 competing systems. 
Even though the gap is not substantial, the two linear approaches to fusion are outperformed by the two non-linear approaches.

 \begin{table}[!t]
	\centering

	\caption{Performance for the ASVspoof 2019 evaluation partition in terms of pooled min t-DCF and pooled EER for top-performing systems (T05, T45, T60 and T24), four different approaches to fusion (boldface) and baseline systems (B1, B2).}
  \setlength\tabcolsep{3.5pt}
	\begin{tabular}{ *{3}{c}}
		\hline
		 System  & min-tDCF& EER	\\ 
	\hline	\hline
	     T05&0.0069&0.22\\
	     \hline
		 T45~\cite{lavrentyeva2019stc}&0.0510&1.86\\
		 
		 \hline
		 {\bfseries GMM fusion} &\textbf{0.0740}&2.92\\
		\hline
		{\bfseries SVM fusion (polynomial kernel)}&{\bfseries 0.0748}&2.92\\
		\hline
		 T60~\cite{chettri2019ensemble}&0.0755&2.64\\
		\hline
		Optimised LFCC (full-band)&0.0904&3.50\\
		 \hline
		{\bfseries Linear fusion} & {\bfseries 0.0911}&3.38\\
		\hline
		 
		T24& 0.0953 &3.45\\ 
		 \hline

        {\bfseries Multinomial logistic regression fusion} & {\bfseries 0.1182}&4.50\\
		\hline
		
		 LFCC:B2~\cite{todisco2019asvspoof}&0.2116&8.09\\
		 \hline
	     CQCC:B1~\cite{todisco2019asvspoof}&0.2366&9.57\\
		\hline
	
	\end{tabular}
	\label{Tab:comparsion results}
	\vspace{-0.15cm}
\end{table}

\section{Conclusions}
The work reported in this paper investigated whether spoofing attacks leave sub-band artefacts that require specific spoofing countermeasures for detection.  In addition, it sought to determine whether 
non-linear fusion approaches offer better potential to combine the scores produced by ensemble of sub-band classifiers. Extending our prior work, we used a high-resolution base front-end that is adapted using a crude center of mass technique to identify 6 different, additional sub-band front-ends, all of which are used with a GMM-based back-end that is relearned for each feature set. 
Fusion was performed with a variety of different techniques, both linear and non-linear.

Excellent results obtained using a high-resolution, full-band classifier alone demonstrate the importance of the front-end.  This finding could be beneficial to other anti-spoofing researchers that use neural networks with standard, low-resolution front-ends.  A switch to high-resolution front-ends may improve performance; even advanced neural network solutions cannot recover information that is already lost, e.g.\ in spectro-temporal decomposition.  Our results also show that sub-band classifiers can detect reliably all attacks in the development data upon which sub-bands classifiers were learned.  Even though evaluation results are far less promising, fusion results still show that the use of sub-band classifiers helps to improve performance beyond what can be achieved with a full-band classifier alone and that non-linear fusion outperforms linear fusion.

Despite its simplicity, our approach outperforms all but two competing challenge systems. Noting that our approach is learned used only training data and not combined training and development data, as was permitted by ASVspoof 2019 rules, noting also that we did not optimise the approach used for sub-band selection nor tackle spectral overlap in any way, this is a particularly satisfactory result. Other experiments for which we do not have the space to report lend further support to our approach.  
Leave-one-out fusion experiments showed the consistent benefit of sub-band classifiers and non-linear fusion. Also, the use of linearly partitioned sub-bands in an otherwise identical setup gave worse performance and show the merit of attack-specific sub-bands, a finding supported by others authors in concurrent work~\cite{tomi2020subband}, albeit for replay spoofing attacks.

Our future work will investigate non-linear probabilistic linear discriminant analysis back-end techniques; linear approaches which assume matching within-class co-variance proved unsuccessful since the within-class co-variance of bona fide and spoofed data is different.  Score normalisation before fusion could also be explored as a means to improve performance using linear fusion.  Another natural extension is to explore the use of high-resolution and sub-band front-ends with neural network based architectures.  The goal of this work would be to see if localised spectral information is being used in the same way and hence to improve upon the interpretability and explainability of complex neural network techniques.

\section{Acknowledgements}
The work was partially supported by the Voice Personae and RESPECT projects, both funded by the French Agence Nationale de la Recherche~(ANR).

\vfill

\balance
\bibliographystyle{IEEEtran}

\begin{thebibliography}{10}
\providecommand{\url}[1]{#1}
\csname url@samestyle\endcsname
\providecommand{\newblock}{\relax}
\providecommand{\bibinfo}[2]{#2}
\providecommand{\BIBentrySTDinterwordspacing}{\spaceskip=0pt\relax}
\providecommand{\BIBentryALTinterwordstretchfactor}{4}
\providecommand{\BIBentryALTinterwordspacing}{\spaceskip=\fontdimen2\font plus
\BIBentryALTinterwordstretchfactor\fontdimen3\font minus
  \fontdimen4\font\relax}
\providecommand{\BIBforeignlanguage}[2]{{%
\expandafter\ifx\csname l@#1\endcsname\relax
\typeout{** WARNING: IEEEtran.bst: No hyphenation pattern has been}%
\typeout{** loaded for the language `#1'. Using the pattern for}%
\typeout{** the default language instead.}%
\else
\language=\csname l@#1\endcsname
\fi
#2}}
\providecommand{\BIBdecl}{\relax}
\BIBdecl

\bibitem{Sahidullah15}
M.~Sahidullah, T.~Kinnunen, and C.~Hanil{\c{c}}i, ``A comparison of features
  for synthetic speech detection,'' in \emph{Proc. INTERSPEECH}, Dresden,
  Germany, 2015, pp. 2087--2091.

\bibitem{sriskandaraja2016investigation}
K.~Sriskandaraja, V.~Sethu, P.~N. Le, and E.~Ambikairajah, ``Investigation of
  sub-band discriminative information between spoofed and genuine speech,'' in
  \emph{Proc. INTERSPEECH}, San Francisco, USA, 2016, pp. 1710--1714.

\bibitem{witkowski2017audio}
M.~Witkowski, S.~Kacprzak, P.~Zelasko, K.~Kowalczyk, and J.~Galka, ``Audio
  replay attack detection using high-frequency features.'' in
  \emph{Interspeech}, 2017, pp. 27--31.

\bibitem{nagarsheth2017replay}
P.~Nagarsheth, E.~Khoury, K.~Patil, and M.~Garland, ``Replay attack detection
  using {DNN} for channel discrimination.'' in \emph{Interspeech}, 2017, pp.
  97--101.

\bibitem{lin2018replay}
L.~Lin, R.~Wang, and Y.~Diqun, ``A replay speech detection algorithm based on
  sub-band analysis,'' in \emph{International Conference on Intelligent
  Information Processing (IIP)}, Nanning, China, 2018, pp. 337--345.

\bibitem{yang2019significance}
J.~Yang, R.~K. Das, and H.~Li, ``Significance of subband features for synthetic
  speech detection,'' \emph{IEEE Transactions on Information Forensics and
  Security}, 2019.

\bibitem{garg2019subband}
S.~Garg, S.~Bhilare, and V.~Kanhangad, ``Subband analysis for performance
  improvement of replay attack detection in speaker verification systems,'' in
  \emph{International Conference on Identity, Security, and Behavior Analysis
  (ISBA)}, 2019, pp. 1--7.

\bibitem{jung2019replay}
J.~Jung, H.~Shim, H.~Heo, and H.~Yu, ``Replay attack detection with
  complementary high-resolution information using end-to-end {DNN} for the
  {ASV}spoof 2019 challenge,'' in \emph{Interspeech}, 2019, pp. 1083--1087.

\bibitem{odyssey2020CQCC}
H.~Tak, J.~Patino, A.~Nautsch, N.~Evans, and M.~Todisco, ``{An explainability
  study of the constant {Q} cepstral coefficient spoofing countermeasure for
  automatic speaker verification},'' \emph{Speaker Odyssey Workshop}, 2020.

\bibitem{ji2017ensemble}
Z.~Ji, Z.~Y. Li, P.~Li, M.~An, S.~Gao, D.~Wu, and F.~Zhao, ``Ensemble learning
  for countermeasure of audio replay spoofing attack in {ASV}spoof2017.'' in
  \emph{INTERSPEECH}, 2017, pp. 87--91.

\bibitem{chettri2019ensemble}
B.~Chettri, D.~Stoller, V.~Morfi, M.~A.~M. Ram{\'\i}rez, E.~Benetos, and B.~L.
  Sturm, ``Ensemble models for spoofing detection in automatic speaker
  verification,'' in \emph{Proc. INTERSPEECH}, Graz, Austria, 2019, pp.
  1118--1112.

\bibitem{Ensemble2020}
J.~Monteiro, J.~Alam, and T.~H. Falk, ``{A}n ensemble based approach for
  generalized detection of spoofing attacks to automatic speaker recognizers,''
  in \emph{ICASSP}, Barcelona, May 4-8, 2020.

\bibitem{todisco2019asvspoof}
M.~Todisco, X.~Wang, V.~Vestman, M.~Sahidullah, H.~Delgado, A.~Nautsch,
  J.~Yamagishi, N.~Evans, T.~Kinnunen, and K.~Lee, ``{ASV}spoof 2019: Future
  horizons in spoofed and fake audio detection,'' in \emph{Proc. INTERSPEECH},
  Graz, Austria, 2019, pp. 1008--1012.

\bibitem{wang2019asvspoof}
X.~Wang, J.~Yamagishi, M.~Todisco, H.~Delgado, A.~Nautsch, N.~Evans,
  M.~Sahidullah, V.~Vestman, T.~Kinnunen, K.~Lee \emph{et~al.}, ``The
  {ASV}spoof 2019 database,'' \emph{arXiv preprint arXiv:1911.01601}, 2019.

\bibitem{ASVspoof19_evalplan}
\BIBentryALTinterwordspacing
{ASVspoof} 2019: the automatic speaker verification spoofing and
  countermeasures challenge evaluation plan. [Online]. Available:
  \url{http://www.asvspoof.org/asvspoof2019/asvspoof2019_evaluation_plan.pdf}
\BIBentrySTDinterwordspacing

\bibitem{kinnunen2018t}
T.~Kinnunen, K.~Lee, H.~Delgado, N.~Evans, M.~Todisco, J.~Sahidullah,
  M.and~Yamagishi, and D.~A. Reynolds, ``{t-DCF: a detection cost function for
  the tandem assessment of spoofing countermeasures and automatic speaker
  verification},'' in \emph{{Proc. Speaker Odyssey Workshop}}, Les Sables
  d'Olonne, France, 2018, pp. 312--319.

\bibitem{kinnunen-tDCF-TASLP}
T.~Kinnunen, H.~Delgado, N.~Evans, K.~A. Lee, V.~Vestman, A.~Nautsch,
  M.~Todisco, X.~Wang, M.~Sahidullah, J.~Yamagishi, and D.~A. Reynolds,
  ``Tandem assessment of spoofing countermeasures and automatic speaker
  verification: Fundamentals,'' \emph{IEEE/ACM Transactions on Audio Speech and
  Language Processing (TASLP)}, 2020, submitted.

\bibitem{brummer2013bosaris}
N.~Br{\"u}mmer and E.~De~Villiers, ``The {bosaris} toolkit: Theory, algorithms
  and code for surviving the new dcf,'' \emph{arXiv preprint arXiv:1304.2865},
  2013.

\bibitem{todisco2016new}
M.~Todisco, H.~Delgado, and N.~Evans, ``A new feature for automatic speaker
  verification anti-spoofing: Constant {Q} cepstral coefficients,'' in
  \emph{Proc. Speaker Odyssey Workshop}, Bilbao, Spain, 2016, pp. 249--252.

\bibitem{todisco2017constant}
------, ``{Constant Q cepstral coefficients: A spoofing countermeasure for
  automatic speaker verification},'' \emph{Computer Speech \& Language},
  vol.~45, pp. 516--535, 2017.

\bibitem{brown1991calculation}
J.~C. Brown, ``{Calculation of a constant Q spectral transform},'' \emph{The
  Journal of the Acoustical Society of America (JASA)}, vol.~89, no.~1, pp.
  425--434, 1991.

\bibitem{schorkhuber2014matlab}
C.~Sch{\"o}rkhuber, A.~Klapuri, N.~Holighaus, and M.~D{\"o}rfler, ``A matlab
  toolbox for efficient perfect reconstruction time-frequency transforms with
  log-frequency resolution,'' in \emph{Proc. Audio Engineering Society
  International Conference on Semantic Audio}, London, UK, 2014.

\bibitem{deller1993discrete}
J.~R. Deller~Jr, J.~G. Proakis, and J.~H. Hansen, \emph{Discrete time
  processing of speech signals}.\hskip 1em plus 0.5em minus 0.4em\relax
  Prentice Hall PTR, 1993.

\bibitem{bhattacharyya1943measure}
A.~Bhattacharyya, ``On a measure of divergence between two statistical
  populations defined by their probability distributions,'' \emph{Bull.
  Calcutta Math. Soc.}, vol.~35, pp. 99--109, 1943.

\bibitem{Feynman:1494701}
\BIBentryALTinterwordspacing
R.~P. Feynman, R.~B. Leighton, and M.~Sands, \emph{{The Feynman lectures on
  physics; New millennium ed.}}\hskip 1em plus 0.5em minus 0.4em\relax New
  York, NY: Basic Books, 2010, originally published 1963-1965. [Online].
  Available: \url{https://cds.cern.ch/record/1494701}
\BIBentrySTDinterwordspacing

\bibitem{cortes1995support}
C.~Cortes and V.~Vapnik, ``Support-vector networks,'' \emph{Machine learning},
  vol.~20, no.~3, pp. 273--297, 1995.

\bibitem{bishop:2006:PRML}
C.~M. Bishop, \emph{Pattern Recognition and Machine Learning}.\hskip 1em plus
  0.5em minus 0.4em\relax Springer, 2006.

\bibitem{todisco2018integrated}
M.~Todisco, H.~Delgado, K.~Lee, M.~Sahidullah, N.~Evans, T.~Kinnunen, and
  J.~Yamagishi, ``Integrated presentation attack detection and automatic
  speaker verification: Common features and gaussian back-end fusion,'' in
  \emph{Proc. INTERSPEECH}, Hyderabadd, India, 2018, pp. 77--81.

\bibitem{lavrentyeva2019stc}
G.~Lavrentyeva, S.~Novoselov, A.~Tseren, M.~Volkova, A.~Gorlanov, and
  A.~Kozlov, ``{STC} antispoofing systems for the {ASV}spoof2019 challenge,''
  in \emph{Proc. INTERSPEECH}, Graz, Austria, 2019, pp. 1033--1037.

\bibitem{tomi2020subband}
B.~Chettri, T.~Kinnunen, and E.~Benetos, ``{Subband modeling for spoofing
  detection in automatic speaker verification},'' \emph{Speaker Odyssey
  Workshop}, 2020.

\end{thebibliography}

\balance

\end{document}